\documentclass[twoside]{article}
\usepackage{fleqn}
\usepackage{espcrc2}
\usepackage{epsf}
\usepackage{graphics}

\newcommand{\be} {\begin{equation}}
\newcommand{\ee} {\end{equation}}
\newcommand{\bea}{\begin{eqnarray}}
\newcommand{\eea}{\end{eqnarray}}

\newcommand{\bi}{\bibitem}

\newcommand{\inc}[1]{\resizebox{7.0cm}{!}{
\rotatebox{-90}{\includegraphics{#1}}}}

\title{\vspace{-4.0cm} 
\begin{flushright}
\vspace{-0.3cm}
{\normalsize\tt BNL-HET-00/41}\\
\end{flushright}
\vspace*{3.0cm}
Non-perturbative renormalisation with domain wall fermions
}
\author{C.~Dawson 
\address{Physics Department, Brookhaven
National Laboratory, Upton, NY 11973-5000, USA}
\thanks{Supported by the U.S.~Department of Energy
under contract number DE-AC02-98CH10886 
and by the RIKEN-BNL Research Center.
Work done in collaboration with T. Blum,
N. Christ, C. Cristian, G. Fleming, X. Liao, G. Liu,
C. Malureanu, R. Mawhinney, S. Ohta,
A. Soni, M. Wingate, P. Vranas, L. Wu, and Y. Zhestkov .}\\ 
RBC Collaboration
}
\begin{document}
\begin{abstract}  
We present results from a study of the renormalisation of both quark bilinear
and four-quark operators for the domain wall fermion action, using the
non-perturbative renormalisation technique of the Rome-Southampton
group. These results are from a quenched simulation, on a $16^3 \times 32$
lattice, with $\beta=6.0$ and $L_s=16$.
\end{abstract}
\maketitle
\section{Introduction}

Domain wall fermions \cite{ref:shamir} are increasingly being used for many
studies of physically interesting quantities.  It is therefore important to
have reliable values for the renormalisation coefficients of the operators
needed  in these calculations. Here we will discuss the application of the
nonperturbative renormalisation technique of the Rome-Southampton group
\cite{ref:gen_npr}, the Regularisation Independant (RI) scheme, to quark bilinears and four-quark operators
using the domain wall fermion action. Rather than concentrating on the
numerical values for the renormalisation coefficients, the emphasis will be on
highlighting tests of the chiral properties of domain wall fermions and on
interesting aspects of the non-perturbative renormalisation technique.

All the results presented are from a study using a lattice with dimensions
$16^3 \times 32$, an extent in the fifth dimension of $16$ points, a 
domain wall height ($M_5$) of $1.8$ and a Wilson gauge action with $\beta=6.0$.

\vspace{-0.1cm}
\section{The RI scheme}

A full description of the RI scheme may be
found in \cite{ref:gen_npr}, but the relevant
details will be quickly summarised here.

The matrix element of the operator of interest, between external quark states,
at high momenta, in a fixed gauge (in this case Landau gauge was used) is
calculated. The external legs of this matrix element are then amputated. A
renormalisation condition may be defined by requiring that, for the
renormalised operator at given scale, a chosen spin and colour projection of
this quantity is equal to its free case value.

\vspace{-0.1cm}

\section{Quark Bilinears}

For the case of flavour non-singlet fermion bilinear operators,       
we write
\be
\left[ \overline{u} \Gamma_i d \right]_{\mathrm{ren}} 
= Z_{\Gamma} \left[ \overline{u} \Gamma_i d \right]
\ee
with 
$
\Gamma_i = \left\{ 
1 , \gamma_\mu , \gamma_5, \gamma_\mu \gamma_5 , \sigma_{\mu \nu}
\right\}$,
where $i$ represents whatever indices the gamma matrices have.
Taking projections of the amputated matrix elements with the 
same gamma matrices as in the definition of the operator, normalised
such that this trace is unity in the free theory, gives the renormalisation
condition
\be
1
= \left. \frac{Z_{\Gamma}}{Z_q} \Lambda_{\Gamma_i} \right|_{ m_{\mathrm{ren}} = 0}
\ee
with
\be
\Lambda_{\Gamma}
=
\frac{1}{\mathrm{Tr} \left[
\Gamma_i \Gamma_i
\right]} \mathrm{Tr}
\left[\Gamma_i
\langle \left[
\overline{u} \Gamma_i d \right]  u(p) \overline{d}(p) \rangle_{AMP}
\right] \, .
\ee

\begin{figure}[!htb]
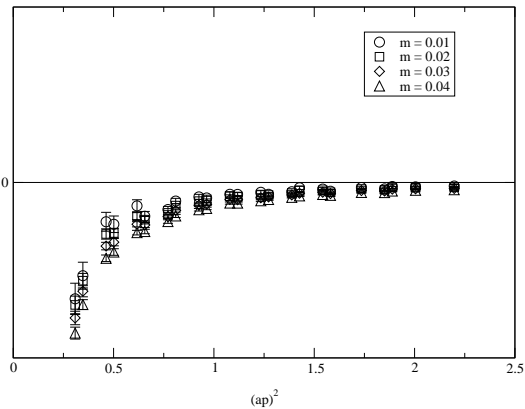

\begin{center}
\inc{fig/Fig1.eps}
\vspace{-0.9cm}
\caption{A plot of  $\Lambda_A - \Lambda_V$ versus $(ap)^2$, showing
that there is no significant difference between $Z_A$ and $Z_V$.}
\vspace{-0.9cm}
\label{fig:zav}
\end{center}
\end{figure}

\begin{figure}[!htb]
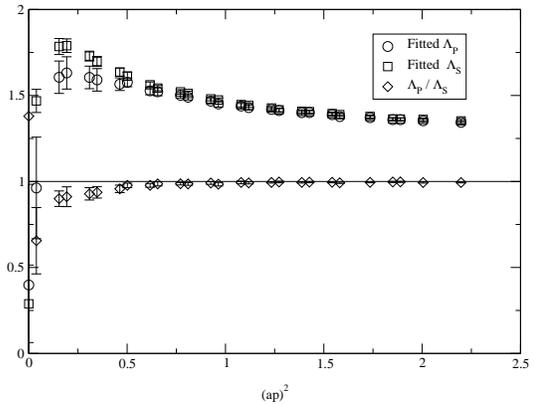

\begin{center}
\inc{fig/Fig2.eps}
\vspace{-0.9cm}
\caption{A plot of $\Lambda_S$, $\Lambda_P$ and their ratio, after the
mass poles have been subtracted, showing that the any splitting between
$Z_S$ and $Z_P$ is smaller that the resolution of this study.}
\vspace{-0.9cm}
\label{fig:zsp}
\end{center}
\end{figure}

Invariance of the action under axial flavour transformations requires that $
Z_A=Z_V$ and $Z_S=Z_P$. The extent to which these relations hold for
domain wall fermions is therefore an excellent test of their chiral
properties. Figure \ref{fig:zav} shows data for $\Lambda_A - \Lambda_V$ versus
$(ap)^2$ for several values of the input mass, $m_f$. While there is a small
difference between $\Lambda_A$ and $\Lambda_V$ at low momenta, this difference
becomes smaller as either the momentum becomes larger or the mass becomes
smaller, suggesting that the bulk of this difference is due to the effects of
chiral symmetry being spontaneously broken by the vacuum and softly broken by
the mass respectively. Within the statistics there is therefore no signal for
a difference between $Z_A$ and $Z_V$.
Figure \ref{fig:zsp} shows $\Lambda_S$, $\Lambda_P$ and their ratio after
the mass poles \cite{ref:gen_npr,ref:masspoles} have been subtracted by fitting the 
mass dependence. Again, within the accuracy of the method there is
no signal for a splitting of $Z_S$ and $Z_P$.
\vspace{-0.1cm}
\section{Four Quark Operators}

The renormalisation of four-quark operators is much more complicated than that
of the quark bilinears considered above. Mixing both among the
four-quark operators and between these operators and lower dimensional
operators must be considered. However, these operators are of much interest
phenomenologically, in particular because of their relevance for the
calculation of $B_K$ and matrix elements of the $\Delta S=1$ Hamiltonian
\cite{ref:matrixel}.

\subsection{$\Delta S=2$ Hamiltonian}

The $\Delta S=2$ Hamiltonian,
which is relevant for calculating $B_K$, is one of the simplest
four-quark renormalisation problems, as mixing with lower dimensional operators 
is not possible. 
It consists of the operators $ O_{AB} = \overline{s} \, \Gamma_A \,
d \, \overline{s} \,
\Gamma_B \, d$, where $\Gamma_A$ and $\Gamma_B$ are  arbitrary
gamma matrices except for the condition that there be no free indices.

As with the bilinears, renormalisation conditions may be placed on these
operators by considering their amputated matrix elements between external
quark states. As mixing between four-quark operators is possible, a
single renormalisation condition no longer suffices; however, by varying the
projectors used, all the renormalisation coefficients can be
fixed \cite{ref:fourquark,ref:oldme}.

To completely specify the renormalisation condition,
the momentum configuration for the external quark states must also be chosen.
For such calculations to be useful, however,
we are required to use the same momenta configuration used in existing
perturbative
matching calculations, which take all the momenta to be equal.

\begin{figure}[!htb]
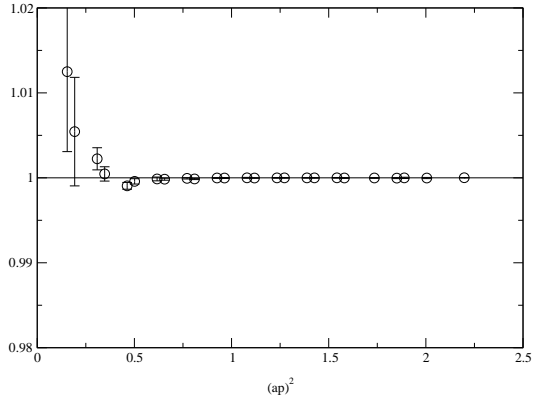

\begin{center}
\inc{fig/Fig3.eps}
\vspace{-0.9cm}
\caption{The ratio of the parity positive and parity negative renormalisation
constants for the $\Delta S=2$ Hamiltonian. No evidence of explicit chiral symmetry
breaking can be seen.}
\vspace{-0.9cm}
\label{fig:parity}
\end{center}
\end{figure}

The renormalisation structure of these operators is strongly
constrained by chiral symmetry \cite{ref:fourquark}. 
Previously presented results \cite{ref:oldme} have confirmed
that the parity conserving part of the operator
$
\overline{s} \gamma_\mu \left( 1 - \gamma_5 \right) d 
\, \overline{s} \gamma_\mu \left( 1 - \gamma_5 \right) d
$
does not significantly mix with any of the four 
wrong chirality operators it could possibly mix with.
Chiral symmetry also predicts  that
the renormalisation factor for the parity conserving and parity breaking 
components of the operator are equal. The ratio of these factors
is shown in Figure \ref{fig:parity}, with no evidence for chiral
symmetry breaking visible for moderately high momenta.

\begin{figure}[!htb]
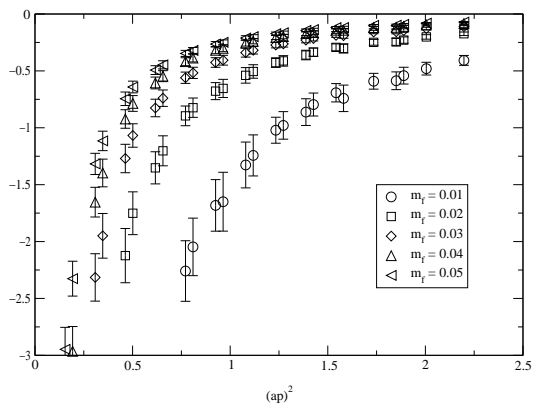

\begin{center}
\inc{fig/Fig4.eps}
\vspace{-0.9cm}
\caption{The element of the mixing matrix connecting $O^+$ and $O^-$, which
chiral symmetry constrains to be zero.}
\vspace{-0.9cm}
\label{fig:masspole}
\end{center}
\end{figure}

Another prediction of chiral symmetry is that the operators
$O^{\pm} = \overline{s}d\,\overline{s}d \pm \overline{s}\gamma_5 d\,\overline{s}
\gamma_5 d$ do not mix with one another.  
The element of the inverse mixing matrix resulting
from a naive application of  the RI scheme renormalisation conditions
to this case is shown in Figure \ref{fig:masspole}
and, as can be seen, this is significantly non-zero. The mass
behaviour is also very distinctive, with the signal being further away from zero
the smaller $m_f$ is. This is not, however, a signal for a large mixing
between these two operators, but an interesting systematic effect 
in the matrix elements we are calculating.

A simple understanding of this effect can be gained by noting that
while we are interested in the matrix elements of operators between quark 
states, we are still evaluating this matrix element in a theory
with propagating pseudo-goldstone bosons. We may approximately
represent the effects of the pseudo-goldstone
bosons by an effective interaction of the form 
\be
C \left\{ \overline{\pi}\, \overline{d} \gamma_5 u
+ 
\overline{u} \gamma_5 d \, \pi 
\right\} \, ,
\label{eq:pole}
\ee
with each such interaction between the quark states and the
pseudo-goldstone bosons giving rise to a propagator of the form 
\be
\frac{1}{ ( p - q )^2 + M_\pi^2 } \, ,
\label{prop}
\ee
where $p$ and $q$ are the momenta on the incoming and outgoing
quarks. 

In the case we are considering, the problem is that all the
external quark states have equal momenta and 
any interaction between them through the vertex in
Equation~\ref{eq:pole} will give a contribution
to the matrix element $\propto 1/M_\pi^2 \equiv
1/m_{\mathrm{quark}}$.

\begin{figure}[!htb]
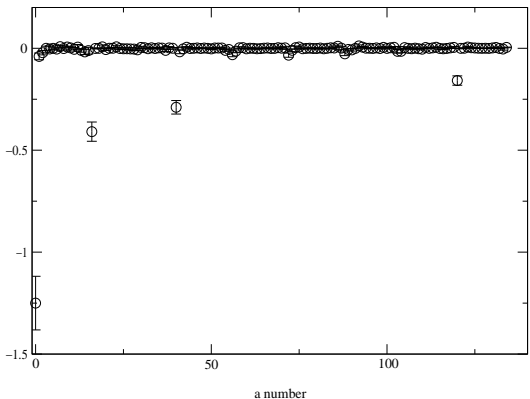

\inc{fig/Fig5.eps}
\vspace{-0.9cm}
\caption{The same quantity as shown in Figure \ref{fig:masspole} 
except that the momenta for the external states is such that the 
pseudo-goldstone boson pole is suppressed for all but four points
on this graph.}
\label{fig:nomasspole}
\vspace{-0.9cm}
\end{figure}

Figure \ref{fig:nomasspole} shows the same quantity as Figure
\ref{fig:masspole}, for $m_f=0.02$, except that the momenta of the incoming
external states, $p$, and outgoing states, $q$, can be different.  The data on
the graph is such that for each point $p^2=q^2$ with four distinct blocks of
values of $p^2$ being shown, while $(p-q)^2$ varies between the different
points.  For all of the points except the four that are distinctly lower, $p
\ne q$, while for those four points $(p-q)^2=0$.  The points for which
$(p-q)^2 \ne 0$ are clearly suppressed as is expected from
Equation~\ref{prop}.

\subsection{$\Delta S = 1$ Hamiltonian}

The mixing problem for the $\Delta S = 1$ Hamiltonian is much more complicated
that for the $\Delta S = 2$ Hamiltonian, both in the number of four-quark
operators that may mix and because of the presence of mixing with lower dimensional
operators. The latter must be subtracted before the standard RI scheme
conditions are applied, and it is the feasibility of this subtraction 
for domain wall fermions that we will touch on here.

The fact that the RI conditions are applied in a fixed gauge  and for
off-shell quark states, makes the problem of mixing with lower dimensional
operators very difficult \cite{ref:me}, with many operators that
must be subtracted. However, this number is greatly reduced if chiral symmetry
is respected. Also, when calculating the renormalisation coefficients in the
context of a practical study, perturbation theory must be relied upon for the
running of the effective Hamiltonian and scheme matching. Operators that do not
appear in these calculations (because they are higher order in perturbation
theory) need not be considered unless they are power divergent in the lattice
spacing.

\begin{figure}[!htb]
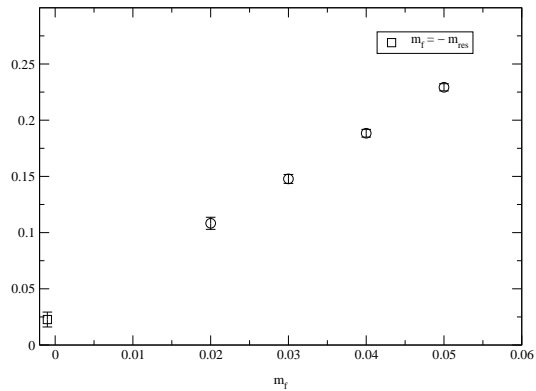

\begin{center}
\inc{fig/Fig6.eps}
\vspace{-0.9cm}
\caption{The coefficient of $\overline{s}d$ for its subtraction from
$O_6$. The strong mass dependence is a sign that, to a good approximation, 
chiral symmetry is 
respected.}
\vspace{-0.9cm}
\label{fig:sub6}
\end{center}
\end{figure}

As it is chiral symmetry that makes performing this subtraction feasible, here
we will give one example of a subtraction coefficient and its chiral
properties. The $\Delta S=1$ operator $O_6 = (\bar{s}_\alpha d_\beta)_{L} \,
\sum_q(\bar{q}_\beta q_\alpha)_{R}$ may mix with the lower dimensional
operator $\overline{s}d$.  If chiral symmetry is not respected, then the
needed subtraction coefficient is cubically power divergent in the lattice
spacing with only subleading mass dependence. Chiral symmetry, however,
constrains the leading behaviour of this coefficient to be $\propto
(1/a^2)(m_s+m_d)$.  Figure \ref{fig:sub6} shows the behaviour of this
subtraction coefficient versus $m_f$, for a single momentum value. While this
does not extrapolate exactly to zero at the same point as the renormalised
mass ($m=-m_{\mathrm{res}}$\cite{ref:rbc_spect}), using a linear
extrapolation, the constant piece is small and is of the order of potential
subleading effects.
\vspace{-0.1cm}
\section{Conclusions}
In conclusion, for the lattice parameters used in this study, domain
wall fermions have excellent chiral properties.
Also, the non-perturbative
renormalisation technique of the Rome-Southampton group seems to work
well as long as one is careful when choosing the
momentum configuration used.

\end{document}